\documentclass[aps,twocolumn,preprintnumbers,floatfix,nofootinbib]{revtex4-1}
\usepackage[utf8]{inputenc}
\usepackage[dvips]{graphicx}
\usepackage{xcolor}
\usepackage{color}
\usepackage{relsize}
\usepackage{graphics}
\usepackage{epstopdf}
\usepackage{hyperref}
\usepackage{mathrsfs}
\usepackage{ragged2e}
\usepackage{amssymb}
\usepackage{subcaption}
\usepackage[normalem]{ ulem }
\usepackage{amsthm}
\usepackage{amsmath}
\usepackage{hyperref}
\usepackage{cancel}
\newcommand{\be}{\begin{equation}}
\newcommand{\ee}{\end{equation}}
\newcommand{\bea}{\begin{eqnarray}}
\newcommand{\eea}{\end{eqnarray}}

\newcommand{\dd}{\text{d}}

\newcommand{\F}{{\cal F}}
\newcommand{\V}{{\cal V}}

\newcommand{\Vone}{{\cal V}_{\mbox{\scriptsize 1-loop}}}
\usepackage{tikz}
\usepackage{pgfplots}
\usetikzlibrary{trees}
\usetikzlibrary{decorations.pathmorphing}
\usetikzlibrary{decorations.markings}
\usetikzlibrary{decorations.text}
   \definecolor{greeen}{rgb}{0.03,0.54,0.23}
\definecolor{test}{rgb}{0.03,0.74,0.33}
\definecolor{viol}{rgb}{0.44,0,0.94}
\definecolor{or}{rgb}{0.9,0.6,0}
\tikzset{
    photon/.style={decorate, decoration={snake, amplitude=2pt}, draw=green},
    photon2/.style={decorate, decoration={snake, amplitude=2pt}, draw=viol},
    dark/.style={draw=greeen, postaction={decorate},
        decoration={markings,mark=at position .5 with {\arrow[draw=greeen]{>}}}},
antidark/.style={draw=greeen, postaction={decorate},
        decoration={markings,mark=at position .5 with {\arrow[draw=greeen]{<}}}},
electron/.style={draw=viol, postaction={decorate},
        decoration={markings,mark=at position .5 with {\arrow[draw=viol]{>}}}},
        antielectron/.style={draw=viol, postaction={decorate},
        decoration={markings,mark=at position .5 with {\arrow[draw=viol]{<}}}},
        neutrino/.style={draw=orange, postaction={decorate},
        decoration={markings,mark=at position .5 with {\arrow[draw=orange]{>}}}},
        antineutrino/.style={draw=orange, postaction={decorate},
        decoration={markings,mark=at position .5 with {\arrow[draw=orange]{<}}}},
gluon/.style={decorate, draw=or,
        decoration={coil,amplitude=4pt, segment length=4pt}},
  ZZ/.style={decorate, decoration={snake}, draw=yellow},    
  scalar1/.style={decorate, dashed, draw=cyan},  
  scalar0/.style={decorate, dashed, draw=greeen},  
   scalar2/.style={decorate, dashed, draw=viol}, 
 pseudoscalar/.style={decorate, dashed, draw=purple},  
   }

\usetikzlibrary{decorations, decorations.markings, decorations.pathmorphing, arrows, graphs, shapes.geometric, snakes}
\usetikzlibrary{shadings}

\begin{document}
\sloppy  

\preprint{}

\begin{flushleft}December 2019, CPHT-RR065.112019\end{flushleft}

\title{Spontaneous Freeze Out of Dark Matter\\ From an Early Thermal Phase Transition}
\author{Lucien Heurtier$^{a}$}
\email{heurtier@email.arizona.edu}
\author{Herv\'e Partouche$^{b}$}
\email{herve.partouche@polytechnique.edu}
\affiliation{\vspace{0.3cm}
${}^a$ 
Department of Physics, University of Arizona, Tucson, AZ   85721\\
${}^b$ 
CPHT, CNRS, Ecole polytechnique, IP Paris, F-91128 Palaiseau, France}

\begin{abstract} 
We propose a new paradigm for the thermal production of dark matter in the early universe, in which dark-matter particles acquire their mass and freeze out {\em spontaneously} from the thermal bath after a dark phase transition takes place. The decoupling arises because the dark-matter particles become suddenly non-relativistic and not because of any decay channel becoming kinematically close. We propose a minimal scenario in which a scalar and a fermionic dark-matter are in thermal equilibrium with the Standard-Model bath. We compute the finite temperature corrections to the scalar potential and identify a region of the parameter space where the fermionic dark-matter mass spontaneously jumps over the temperature when the dark phase transition happens. We explore the phenomenological implications of such a model 
in simple cases and show that the annihilation cross section of dark-matter particles has to be larger by more than one order of magnitude as compared to the usual constant-mass WIMP scenario in order to accomodate the correct relic abundance. We show that in the spontaneous freeze out regime a TeV-scale fermionic dark-matter that annihilates into leptons through $s$-wave processes can be  accessible to detection in the near future.
\end{abstract}

\maketitle

\setcounter{equation}{0}


{\section{Introduction}\label{intro}
The nature of dark matter (DM) and the way it is produced in the early universe stand among the biggest puzzles of modern cosmology. The Weakly Interacting Massive Particle (WIMP) paradigm has triggered a lot of attention in the past decades due to its simplicity and to the surprising connection which exists in this framework between the  energy scale necessary to produce the correct DM relic abundance and the electroweak scale \cite{Bergstrom:2000pn,Bertone:2004pz,Steigman:2012nb,Patrignani:2016xqp}. 
By construction, WIMP scenarios rely on two major ingredients, which are  that  $(i)$ the dark-matter particles  thermalize with the Standard-Model (SM) bath at high temperature, and  that $(ii)$ the cross section of annihilation of dark-matter particles into visible states is small enough in order to guarantee that the Freeze Out (FO) mechanism generates a sufficient DM relic abundance in the present universe. As far as theoretical aspects of the model are concerned, the mass of a WIMP candidate cannot be  arbitrarily large. Indeed, too heavy WIMPs, which are required to have a small number density in order to not overclose the universe, need to annihilate efficiently during the Freeze Out, demanding their cross section of annihilation to violate unitarity constraints \cite{Griest:1989wd}. Moreover, vanilla versions of the WIMP involving an electroweak-scale cross section of annihilation of dark-matter particles into nucleons have already been mostly ruled out by experiments from $\mathcal O(10-10^3)~\mathrm{GeV}$ up to the unitarity limit by dark-matter detection experiments and LHC searches\cite{Richard:2014vfa,Carpenter:2012rg,Carpenter:2013xra,Petrov:2013nia,Bell:2012rg,Bell:2015rdw,Birkedal:2004xn,Gershtein:2008bf,Goodman:2010ku,Crivellin:2015wva,Petriello:2008pu,Berlin:2014cfa,Lin:2013sca,Fox:2011pm,Bell:2015sza,Bai:2015nfa,Autran:2015mfa,Gupta:2015lfa,Ghorbani:2016edw,Abercrombie:2015wmb,Escudero:2016gzx,Amole:2015pla,Amole:2016pye,Fu:2016ega,Akerib:2016lao,Aprile:2016swn,Aprile:2017iyp,Cui:2017nnn,Akerib:2016vxi,Aaboud:2016qgg,Aaboud:2016obm,Aaboud:2017dor,Aaboud:2017uak,Aaboud:2017yqz,Aaboud:2017bja,Aaboud:2017rzf,Aaboud:2017phn,Sirunyan:2017onm,Sirunyan:2017hci,Sirunyan:2017hnk,Sirunyan:2017xgm,Sirunyan:2018gka,Sirunyan:2018wcm,Baer:2014eja,Roszkowski:2017nbc,Ackermann:2013yva,Dudas:2013sia,Cirelli:2015gux,Gaskins:2016cha}. In order to avoid direct-detection constraints, one usually has to either seclude dark-matter from the coloured sector, weakening drastically constraints from direct-detection experiments \cite{Fox:2008kb,Bi:2009uj,Cohen:2009fz}, or build models in which the scattering cross-section of DM particles on nucleons is naturally suppressed. Indirect detection constraints can also be avoided by demanding that the cross-section of annihilation of dark-matter particles in the galaxy is velocity suppressed. In this case the flux of cosmic rays produced by the dark sector is reduced and dark matter is effectively invisible in the present universe \cite{Kim:2006af,Hagelin:1984wv,Kim:2008pp,Pospelov:2007mp,Tulin:2013teo,Shelton:2015aqa,Evans:2017kti}.

Alternatives to the WIMP paradigm usually assume that dark-matter particles are produced out of equilibrium, either from the decay of a heavy particle (moduli, inflaton) \cite{Kolb:1998ki,Allahverdi:2018iod} or from the feeble annihilation of particles which are thermalized with the Standard-Model bath \cite{Hall:2009bx,McDonald:2001vt,McDonald:2015ljz,Kaneta:2019zgw,Chu:2013jja,Bhattacharyya:2018evo}. 
In all these different scenarios, the dark-matter mass, and the masses of its decay or annihilation products are assumed to be constant throughout the universe evolution. In the WIMP scenario in particular, the calculation of the dark-matter relic density relies on the evaluation of the annihilation cross-section of dark-matter as a function of masses and couplings, which are assumed to be identical at the time of freeze out and at present time.

In the case of the Standard Model, however, it is known that the whole mass spectrum is temperature dependent. Indeed, the mass of the SM fermions is given by the vacuum expectation value (vev) of the Brout-Englert-Higgs boson. At high temperature, before  the electroweak phase transition, thermal corrections to the Higgs scalar potential  stabilize the latter at the origin $\langle H\rangle=0$ and the Standard Model is essentially composed of pure radiation. At low temperature, when the electroweak phase transition takes place, the $U(1)\times SU(2)$ gauge group is spontaneously broken and SM particles acquire masses proportionally to the Higgs vev $\langle H\rangle\not=0$. Therefore, similarly to the SM particles, the mass spectrum of any thermalized dark sector in which masses orginate from the spontaneous breaking of some UV symmetry group is expected to evolve with the temperature.

The possibility that the dark-matter mass might be a time-dependent quantity was studied in the past in the context of Variable-Mass Particles (VAMPs) in which dark-matter particles interact with a quintessence field \cite{Anderson:1997un,Rosenfeld:2005pw,Franca:2003zg}. Moreover, the effect of thermal corrections to the potential of a dark scalar was also used in the context of the so-called {\em Flip-Flop vev} mechanism \cite{Baker:2017zwx,Baker:2018vos}, {\em super-cool} dark matter \cite{Hambye:2018qjv}, or in the {\em forbidden freeze-in} scenario \cite{Darme:2019wpd}, where a second order phase transition is used to kinematically open or close certain annihilation or decay channels in the early universe. 

In this paper we aim to focus on the very simple case of a thermal fermionic dark matter whose mass is sourced by the vev of a dark scalar. We will study how the thermal corrections arising from the contact of this scalar with thermalized particles drives a thermal, second order phase transition in the dark sector. In particular, we will identify an interesting region of the parameter space where the  phase transition enforces the freeze out to take place before the dark-matter mass reaches its zero-temperature value. We will refer to this possibility  as a {\em Spontaneous Freeze Out} (SFO).  We will  explore in particular to which extent in the SFO case the mass of dark-matter particles at freeze out can differ from its value at present time. We will also show that the annihilation cross section of DM particles into SM states required to obtain the correct DM relic abundance can differ by more than one order of magnitude from the case of a constant-mass WIMP. Therefore, the SFO scenario is more sensitive to DM searches. It is interesting to note that this paradigm was first encountered in the context of string theory \cite{Coudarchet:2018ezq} where it is natural to expect moduli fields to couple to matter states and undergo non-trivial phase transitions.

The paper is organized as follows: In Sec.~\ref{sec:model} we introduce the model on which we shall focus throughout this work. In Sec.~\ref{sec:thermal} we derive the master equations necessary to study the evolution of the mass spectrum with the temperature. In Sec.~\ref{sec:FO} we study the dark-matter freeze out in our model and properly define the paradigm of Spontaneous Freeze Out. In Sec.~\ref{sec:scalar} we study analytically how the decoupling of dark-matter back-reacts on the dynamics of the dark scalar. In Sec.~\ref{sec:SMinteractions}  we finally specify the way dark-matter particles interact with SM states, compute numerically the DM relic density and confront our model to existing direct- and indirect-detection constraints. Our conclusions, comments and perspectives can be found in Sec.~\ref{sec:conclusion}.
}

\section{The Model}\label{sec:model}
 Our benchmark model is based on the tree-level Lagrangian density
\be
\mathcal{L}_{\rm tree}= \mathcal{L}_{\rm SM}+\mathcal{L}_{\rm dark}+\mathcal{L}_{\rm int}\,,
\label{L1}
\ee
where $\mathcal{L}_{\rm SM}$ is the SM contribution. The dark sector contains a fermion $\psi$ and a real scalar field $\phi$ charged under a $\mathbb Z_2$ symmetry ($\phi\to -\phi$). The associated Lagrangian is
\be
\mathcal L_{\rm dark}=i\bar \psi\cancel{\partial}\psi+\frac{1}{2}\partial_\mu \phi\partial^\mu \phi-y \phi \bar\psi\psi -\V_{\rm tree}(\phi)+{\cal L}_{\rm dark}^{\rm c.t.}\,,
\label{lag}
\ee
where ${\cal L}_{\rm dark}^{\rm c.t.}$ stands for counterterms to be tuned suitably later on. The Lagrangian $\mathcal L_{\rm int}$, which is specified in Sec.~\ref{sec:SMinteractions}, contains only interactions between $\psi$, $\phi$ and SM fields. 
 The potential $\V_{\rm tree}(\phi)$ is chosen such that the scalar $\phi$ acquires a non-zero vev and sources a mass term for the dark fermion via the Yukawa interaction,
\be\label{eq:largemass}
m_{\psi}(\langle\phi\rangle)=y \langle\phi\rangle\,.
\ee
Defining the potential as
\be \V_{\rm tree}(\phi)=-\frac{\mu^2}{2} \phi^2+\frac{\lambda}{4!}\phi^4\,,
\ee
where $\mu$ and $\lambda$ are positive, 
the tree-level scalar vev  and the masses of  $\phi$ and $\psi$ are given by
\be
\begin{aligned}
\langle\phi\rangle_{\rm tree}&=\mu\sqrt{\frac{6}{\lambda}}\,,\\
m^{\rm tree}_\phi&=\sqrt 2\, \mu\,,\quad  m^{\rm tree}_\psi=y\sqrt{3\over \lambda}\, m_\phi^{\rm tree}\,.
\end{aligned}
\label{ttree}
\ee

In the following sections, we will show the existence of a region in the parameter space where the dark-matter present in the universe today is composed of fermions $\psi$ only, while the dark scalar particles $\phi$ have decayed into SM states. Even if we consider a Dirac fermion $\psi$ in Eq.~(\ref{lag}), our results will be derived for an arbitrary number $n_F$ of fermionic degrees of freedom, e.g. $n_F=4$ for a Dirac and $n_F=2$ for a Majorana fermion. 


\section{Thermal Effective Potential}
\label{sec:thermal}

From now on, we will assume that the interactions contained in the Lagrangian $\mathcal L_{\rm int}$ are sufficient to maintain the dark sector particles $\psi$ and $\phi$ in thermal equilibrium with the visible sector. At finite temperature, virtual loops of the fields $\psi$ and $\phi$ induce corrections to the scalar potential, which are known to induce a restoration of the vacuum $\mathbb Z_2$ symmetry at high temperature (see e.g. Ref.~\cite{Quiros:1999jp} for a review and references therein). Therefore, when thermal loop corrections dominate over the zero-temperature tree level contribution, perturbation theory breaks down. 

To see this explicitly, let us consider the thermal effective potential at 1-loop, 
\be
\Vone^{\rm th}(T,\phi)=\V_{\rm tree}(\phi)+\V_{\rm CW}(\phi)+\V_{\rm dark}^{\rm c.t.}(\phi)+\F(T,\phi)\,,
\label{Veff1}
\ee
where $\V_{\rm CW}$ is the zero-temperature Coleman-Weinberg contribution and $\F(T,\phi)$ is the free energy. In dimensional regularization, the UV divergence of $\V_{\rm CW}$ can be removed by adjusting counterterms $\V_{\rm dark}^{\rm c.t.}=-\delta_\mu \phi^2+\delta_\lambda \phi^4$, which yield in $\overline{\rm MS}$ scheme the usual result
\be
  \label{CW1}
  \begin{aligned}
\V_{\rm CW}(\phi)+\V_{\rm dark}^{\rm c.t.}(\phi)= &\; \frac{m_0(\phi)^4}{64\pi^2}\left[\log\!\left(\frac{m_0(\phi)^2}{Q^2}\right)-\frac{3}{2}\right]\\
&\!\!\!\!\!\!\! \!\!\!\!\!\!\!-n_F\frac{m_\psi(\phi)^4}{64\pi^2}\left[\log\!\left(\frac{m_\psi(\phi)^2}{Q^2}\right)\!-\frac{3}{2}\right],
\end{aligned}
\ee
where $Q$ is the renormalization energy scale. Hence, $\mu$ and $\lambda$ are running parameters that depend implicitly on $Q$. Moreover, the tree-level mass squared  of $\phi$ is defined as 
\be
m_0(\phi)^2=-\mu^2+{\lambda\over 2}\phi^2\, ,
\ee 
which is negative when $\phi^2<2\mu^2/\lambda$. 

The thermal contribution is the  Helmholtz free energy density of a gaz comprising one bosonic and $n_F$ fermionic degrees of freedom. At 1-loop,  its expression is  
\be
\label{f1}
\F(T,\phi)=\frac{T^4}{2\pi^2}\!\left[J_B\!\left(\frac{m_0(\phi)^2}{T^2}\right)-n_F\, J_F\!\left(\frac{m_\psi(\phi)^2}{T^2}\right)\right],
\ee
where the functions $J_B$ and  $J_F$ are defined as
\be
\label{JBF}
J_{\overset{\scriptstyle B}{F}}\!\left(\frac{m^2}{T^2}\right) =\int_0^{+\infty}du\,  u^2 \log \!\left(1\mp e^{-\sqrt{u^2+m^2/T^2}}\right).
\ee
At hight temperature, these quantities can be expanded as 
\begin{align}
\label{eq:thermalpot}
J_B\!\left(\frac{m^2}{T^2}\right)& =-\frac{\pi^4}{45}+\frac{\pi^2}{12}\frac{m^2}{T^2}-{\pi\over 6}\Big({m^2\over T^2}\Big)^{3\over 2}\nonumber \\
&\;\;\;\;-\frac{1}{32}\frac{m^4}{T^4}\log \frac{m^2}{16\alpha T^2}+\mathcal{O}\!\left(\frac{m^6}{T^6}\right),\\
J_F\!\left(\frac{m^2}{T^2}\right)& =\frac{7\pi^4}{360}-\frac{\pi^2}{24}\frac{m^2}{T^2}-\frac{1}{32}\frac{m^4}{T^4}\log \frac{m^2}{\alpha T^2}+\mathcal{O}\!\left(\frac{m^6}{T^6}\right)\!,\nonumber 
\end{align}
where $\alpha=\pi^2\exp(3/2-2\gamma_{\rm E})$ and $\gamma_{\rm E}$ is the Euler-Mascheroni constant. Note that the  $m_0^4\log m_0^2$ and $m_\psi^4\log m_\psi^2$  terms of the zero-temperature potential and free energy in Eqs~\eqref{CW1},~\eqref{f1} and \eqref{eq:thermalpot} cancel exactly.  
However, comparing the expansions of $J_B$ and $J_F$, one sees that the free energy of the bosonic particles $\phi$  contains an extra source of non-analyticity, namely the term in $(m_0^2)^{3\over 2}$. Therefore, the 1-loop thermal effective potential is complex when $m_0(\phi)^2<0$. In some cases, imaginary parts in effective potentials are associated with physical phase transitions~\cite{Weinberg:1987vp}. However, in the present case, the non-analytic term arises because thermal higher loop corrections that are of the same order of magnitude have been omitted. It turns out that the high-temperature quantum corrections are dominated by the so-called ring (or daisy) diagrams, with an arbitrary number of loops~\cite{Dolan:1973qd, Carrington:1991hz}.  In practice, the resummation of the high-temperature limit of the ring diagrams amounts to adding a contribution to the 1-loop potential~\cite{Delaunay:2007wb},
\be
\V_{\rm eff}^{\rm th}(T,\phi)\equiv \Vone^{\rm th}(T,\phi)+\V_{\rm ring}^{\rm th}(T,\phi)\, , 
\label{vtot}
\ee
where
\be
\V_{\rm ring}^{\rm th}(T,\phi)={T\over 12\pi}\Big[ \big(m_0(\phi)^2\big)^{3\over 2}-\big(m_0(\phi)^2+\Pi_\phi(T)\big)^{3\over 2}\Big].
\ee
In the above expression, the shift of $m_0^2$ is the so-called Debye mass squared
\be
\Pi_\phi(T)= {T^2\over 24}(\lambda+n_Fy^2)\, , 
\ee
which is the dominant temperature contribution of $\partial^2 \F/(\partial\phi)^2$. 
As a result, the pathological contributions $(m_0^2)^{3\over 2}$ of the 1-loop and ring diagrams cancel exactly, while the remaining term $(m_0^2+\Pi_\phi)^{3\over 2}$ in  $\V_{\rm ring}^{\rm th}$ is real.

Even if this is not necessary, we neglect from now on for the sake of simplicity all terms $T^4\times \mathcal{O}(m^6/T^6)$  arising in the high temperature expansions of the 1-loop free energy density $\F$ and ring contribution $\V_{\rm ring}^{\rm th}$. In order to write the thermal effective potential in a suggestive way, we find convenient to set the renormalization scale $Q$ in terms of a critical temperature, 
\be
\begin{aligned}
Q&=\pi e^{-\gamma_{\rm E}} T_c\, ,\\
T_c&={2\sqrt{6}\, \mu\over \sqrt{\lambda+n_F y^2}} \sqrt{1-{\sqrt{6}\over 8\pi}\,\xi+{\log 2\over 8\pi^2}\,\lambda\over 1-{\sqrt{6}\over 4\pi}\,\xi}\,,
\end{aligned}
\ee
where we have defined
\be
\xi\equiv {\lambda\over \sqrt{\lambda+n_Fy^2}}\in \big(\lambda,\sqrt{\lambda} \big)\, .
\ee
Expressing all dependences in $T$  with a new variable $x$, 
\be
x={T_c\over T}\, ,
\label{defi}
\ee
the effective potential at finite temperature reduces to 
\be
\V^{\rm th}_{\rm eff}(x,\phi)=\V_0(x)-\frac{\mu_{\rm eff}(x)^2}{2}\phi^2+\frac{\lambda_{\rm eff}(x)}{4!}\phi^4\,.
\ee
In this expression, $\V_0$ depends only on the temperature, while $\mu^2_{{\rm eff}}$ is an effective mass term,
\be
\begin{aligned}
\mu_{{\rm eff}}(x)^{2}&=\mu^2\Bigg[\!\bigg(1-{\sqrt{6}\over 8\pi}\,\xi+{\log 2\over 8\pi^2}\,\lambda\bigg)\Big(1-{1\over x^2}\Big)\\
&\;\;\;\;\;\;\;\;\;\;\;\;\;\;\;\;\;\;\;\;\;\;\;\;\;\;\;\;\;\;\;\;\;\;\;\;\;\;\;\;-{\lambda\over 16\pi^2}\log x\Bigg],\\
\end{aligned}
\label{mueff}
\ee
and $\lambda_{{\rm eff}}$ is an effective self-coupling,
\be
\begin{aligned}
\lambda_{{\rm eff}}(x)&=\lambda\bigg(1-{3\sqrt{6}\over 8\pi}\,\xi+{3\log 2\over 8\pi^2}\,\lambda\bigg)\\
&\;\;\;\;\;\;\;\;\;\;\;\;\;\;\;\;\;\;\;\;+{3\over 16\pi^2}\big(4n_F y^4-\lambda^2\big)\log x\, .
\end{aligned}
\label{leff}
\ee
The mass term $\mu_{{\rm eff}}(x)^2$ increases from negative values at $x\ll 1$, up to a positive maximum at $x\simeq 4\pi \sqrt{2/\lambda}$, and it vanishes at $x=1$.\footnote{It is also positive between $x=1$ and the extremely high formal value $x\simeq e^{16\pi^2/\lambda}$, which is far above the domain of validity of the thermal expression of the potential.}  Therefore, the field $\phi$ acquires thermal corrections to its mass, and a phase transition takes place at $x=1$, which justifies $T_c$ to be referred to as a critical temperature.  The scalar $\phi$ is  stabilized at zero at higher temperature, while it condenses and follows adiabatically a temperature-dependent vev at lower temperature. We will discuss in Sec.~\ref{sec:SMinteractions} the validity of such a behavior.


\section{Spontaneous Freeze Out of Dark Matter}\label{sec:FO}

In this section, we  explain how the fermionic dark-matter particles $\psi$  can freeze out, while the dark-scalar ones remain in thermal equilibrium with the SM bath. However, this will be fully justified in Sect.~\ref{sec:SMinteractions}, where we present specific examples of interactions $\mathcal{L}_{\rm int}$ between dark-sector and SM fields.

After the dark-matter particles $\psi$ freeze-out at $T=T_{\rm FO}$, only $\phi$ remains eventually thermalized with the SM. The expression of the effective potential we considered so far is therefore valid up to $x_{\rm FO}=T_c/T_{\rm FO}$. The vev of $\phi$ reads\footnote{If at reheating temperature, $x=x_{\rm RH}$, the effective coupling is negative, $\lambda_{{\rm eff}}(x_{\rm RH})<0$, the vev $\langle \phi\rangle=0$ corresponds to a local minimum of the potential, which becomes global as $x$ approaches 1 from below.  }
\be
\begin{aligned}
x\leqslant1 &:\quad\langle\phi\rangle = 0\,,\\
1\leqslant x\leqslant x_{\rm FO}  &:\quad\langle\phi\rangle = \mu_{{\rm eff}}(x)\sqrt{\frac{6}{\lambda_{\rm eff}(x)}}\,,
\end{aligned}
\label{vevphi}
\ee
and the dark-scalar mass evolves accordingly like
\be
\begin{aligned}
x\leqslant1 &:\quad m_{\phi}(x) = |\mu_{{\rm eff}}(x)|\,,\\
1\leqslant x\leqslant x_{\rm FO}  &:\quad m_{\phi}(x) = \sqrt{2} \,\mu_{\rm eff}(x)\,.
\end{aligned}
\label{mx}
\ee
As a result, the fermionic dark-matter mass becomes effectively a function of the temperature,
\be
\begin{aligned}
x\leqslant1 &:\quad m_{\psi}(x) = 0\, , \\
1\leqslant x\leqslant x_{\rm FO} &:\quad m_{\psi}(x) =  y\sqrt{\frac{3}{\lambda_{\rm eff}(x)}}\, m_{\phi}(x)\,.
\end{aligned}
\label{masspsi}
\ee
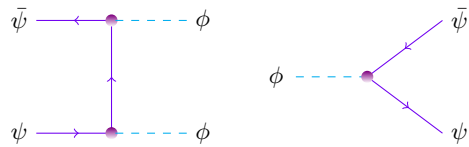
\begin{figure}
\begin{tikzpicture}

        \draw[electron](-2.4,-0.75)--(-2.4,0.75);
      \draw[electron](-2.4,0.75)--(-3.4,0.75)node[left]{$\bar\psi$};
      \draw[antielectron](-2.4,-0.75)--(-3.4,-0.75)node[left]{$ \psi$};
        \shade[top color=violet, bottom color=white]
 (-2.4,0.75)circle(0.08);
         \shade[top color=violet, bottom color=white]
 (-2.4,-0.75)circle(0.08);

  \draw[scalar1](-1.4,0.75)node[right]{$\phi$}--(-2.4,0.75);
    \draw[scalar1](-1.4,-0.75)node[right]{$\phi$}--(-2.4,-0.75);

  \draw[electron](2,0.75)node[right]{$\bar\psi$}--(1,0);
  \draw[antielectron](2,-0.75)node[right]{$\psi$}--(1,0);
  \draw[scalar1](1,0)--(0,0)node[left]{$\phi$};
           \shade[top color=violet, bottom color=white]
 (1,0)circle(0.08);
\end{tikzpicture}
\caption{\label{fig:darkinteractions}\footnotesize Annihilation and decay processes contributing to thermal equilibrium in the dark sector.}
\end{figure}
\begin{figure}
\begin{tikzpicture}
         \draw[electron](-2.4,0)--(-3.4,0.75)node[left]{$\bar\psi$};
      \draw[antielectron](-2.4,0)--(-3.4,-0.75)node[left]{$\psi$};
      \draw[dark](-1.4,0.75)node[right]{$\bar f_{\rm SM}$}--(-2.4,0);
  \draw[antidark](-1.4,-0.7)node[right]{$f_{\rm SM}$}--(-2.4,0);
  \shade[top color=greeen, bottom color=white]
 (-2.4,0)circle(0.08);
      
      \draw[dark](2,0.75)node[right]{$\bar  f_{\rm SM}$}--(1,0);
  \draw[antidark](2,-0.7)node[right]{$f_{\rm SM}$}--(1,0);
  \draw[scalar1](1,0)--(0,0)node[left]{$\phi$};
  \shade[top color=greeen, bottom color=white]
 (1,0)circle(0.08);

\end{tikzpicture}
\caption{\label{fig:SMinteractions}\footnotesize Examples of annihilation and decay processes arising from the interaction Lagrangian $\mathcal L_{\rm int}$ (see Sec.~\ref{sec:SMinteractions}) which may contribute to the thermal equilibrium between $\psi$, $\phi$ and SM particles.}
\end{figure}
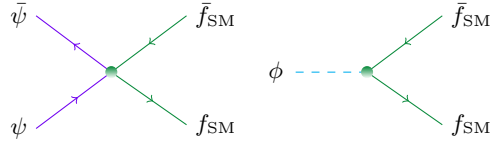
The mechanism which maintains dark-sector particles in thermal equilibrium before they freeze-out can be of different natures. First $\psi$ remains thermalized with the SM  as long as 
\be\label{eq:thermalequilibriumconditionSM}
(i)\; \; H<n_\psi\langle \sigma_{{\rm SM}\leftrightarrow\psi\bar\psi} \,v\rangle\,,
\ee
where  $\langle \sigma_{{\rm SM}\leftrightarrow\psi\bar\psi} \,v\rangle$ denotes the thermally averaged cross-section of annihiliation of $\psi$ into SM fields, $n_\psi$ denotes its number density and $H$ is the Hubble parameter. Second, dark-matter particles can remain in thermal equilibrium with the dark scalar $\phi$, given that
\be\label{eq:thermalequilibriumconditionDark}
(ii)\;\; H<\Gamma_{\phi\to \psi\bar\psi}\quad \mbox{or}\quad H<n_\psi\langle \sigma_{\phi\phi\leftrightarrow\psi\bar\psi} \,v\rangle\,,
\ee
where $\Gamma_{\phi\to \psi\bar\psi}$ stands for the width of the (inverse) decay process $\phi\leftrightarrow\psi+\bar\psi$ and $\langle \sigma_{\phi\phi\leftrightarrow\psi\bar\psi} \,v\rangle$ denotes the annihilation cross section of $\psi$ through a $t$-channel  process (see Fig.~\ref{fig:darkinteractions}).
Finally, $\phi$ can thermalize with the SM as long as either the conditions $(i)$ and $(ii)$ are guaranteed simultaneously, or because interactions between $\phi$ and SM fields are present and such that
\be\label{eq:thermalequilibriumconditionPhi}
(iii)\; \; H<n_\phi\langle \sigma_{{\rm SM}\leftrightarrow\phi\phi} \,v\rangle\quad \mbox{or}\quad H<\Gamma_{\phi\to \mathrm{SM}}\,,
\ee
where $n_\phi$ is the number density of $\phi$, while $\langle \sigma_{\rm SM \leftrightarrow\phi\phi} \,v\rangle$ and $\Gamma_{\phi\to\rm SM}$ are the annihilation cross section and decay width.
In this last case $(iii)$, $\psi$ does not necessarily  have to interact with SM particles in order to thermalize with them, provided the condition $(ii)$ is fulfilled. We depict the different Feynman diagrams which can lead to such interactions in Figs~\ref{fig:darkinteractions} and~\ref{fig:SMinteractions}.

In Sec.~\ref{sec:SMinteractions} we will explicitely introduce interactions between the dark sector and the SM. We will in particular focus on the simple case where
\begin{equation}\label{eq:SMdominate}
\langle \sigma_{{\rm SM}\leftrightarrow\psi\bar\psi} \,v\rangle\gg\langle \sigma_{\phi\phi\leftrightarrow\psi\bar\psi} \,v\rangle\,,
\end{equation}
such that the freeze-out of dark matter is only driven by the contact interaction between dark-matter and SM particles, and not by the interactions taking place within the dark sector. We will also work in a regime in which the dark scalar $\phi$ remains in thermal equilibrium with the SM while the dark-matter freeze out takes place. Therefore the condition $(i)$ will remain true until $x=x_{\rm FO}$. On the scalar side, the condition $(ii)$ will be satisfied when $x<1$ because dark-matter particles are massless and the inverse-decay process is kinematically allowed, whereas the condition $(iii)$ will be satisfied at later time, as we will discuss in Sec.~\ref{sec:SMinteractions}.

In principle, alternative scenarios, violating the condition of Eq.~\eqref{eq:SMdominate}, in which $\psi$ remains in equilibrium with $\phi$ before freezing out could be perfectly viable. This dark equilibrium could take place either together with the SM (if the $\phi\leftrightarrow SM$ interaction is strong enough) or in a dark equilibrium secluded from the SM bath (see e.g. Refs.~\cite{Heurtier:2019eou,Berlin:2016gtr,Cirelli:2018iax}). This possibility would however suppress the interaction between DM and SM fields and be therefore more challenging to detect experimentally.

\subsection*{The Spontaneous Freeze-Out Regime}

To proceed, let us define the ratio $\kappa=m_{\psi}(x_{\rm FO})/T_{\rm FO}$. For $x_{\rm FO}>1$, we obtain from this definition the following expression,  
\be
\begin{aligned}
x_{\rm FO}^2&= 1+\left({4\kappa^2\!\left[\lambda+{3\over 16\pi^2}(4n_F y^4-\lambda^2)\log x_{\rm FO}\right]\over y^2(\lambda+n_Fy^2)}\right.\\
&\qquad \;\;\;\quad +\left.{\lambda\over 16\pi^2}\log x_{\rm FO}\right)\big(1+{\cal O}(\xi)\big),
\end{aligned}
\label{ux}
\ee
which is valid in perturbative regime.
In practice, the value of $\kappa$ depends on the explicit temperature dependency of the DM annihilation cross section into SM particles and is well-known to be of order $\mathcal O(20-30)$ for masses in the GeV-TeV range. It is in Sec.~\ref{sec:SMinteractions} that we will numerically evaluate the freeze-out temperature, using explicit examples of interactions between $\psi$ and SM particles. From Eq.~(\ref{ux}), we observe that different regions of the parameter space can be distinguished.  For a large scalar self-interaction, as compared to the Yukawa interaction, we have   
\be
\lambda\gg n_Fy^2\quad \Longrightarrow\quad x_{\rm FO}\simeq  {\cal O}\Big({2\kappa\over y}\Big)\gg \kappa\,. 
\ee
However, in the reversed case,  we obtain 
\be\label{eq:xFO}
\lambda\ll n_Fy^2\ \Longrightarrow\ x_{\rm FO}\simeq \left[1+\kappa^2\Big({4 \lambda\over n_F y^4}+{3\over \pi^2}\log x_{\rm FO}\Big)\right]^{1/2}\!\!\!,
\ee
which allows $x_{\rm FO}$ to be relatively smaller than $\kappa$ as long as $\lambda< y^4$. We can therefore identify two different regimes: 
\begin{itemize}
\item  When $x_{\rm FO}\gg \kappa$, the temperature at which dark-matter freezes out is much lower than  the critical one, $T_{\rm FO}\ll T_c$. At such a low temperature, the masses $m_\phi(x_{\rm FO})$ and $m_\psi(x_{\rm FO})$ (see Eqs~\eqref{mx},~(\ref{masspsi}) and~(\ref{mueff}),~(\ref{leff})) are close to their tree-level values, 
\be
\begin{aligned}
\qquad 1\ll x\leqslant x_{\rm FO}  :\quad &m_{\phi}(x) \simeq m_{\phi}^{\rm tree} \,, \\
&m_{\psi}(x) \simeq m_\psi^{\rm tree}\, .
\end{aligned}
\ee
FIG.~\ref{fig:a} 
\begin{figure*}
\begin{subfigure}[b]{0.45\linewidth}
\caption{\label{fig:a}}
\includegraphics[width=\linewidth]{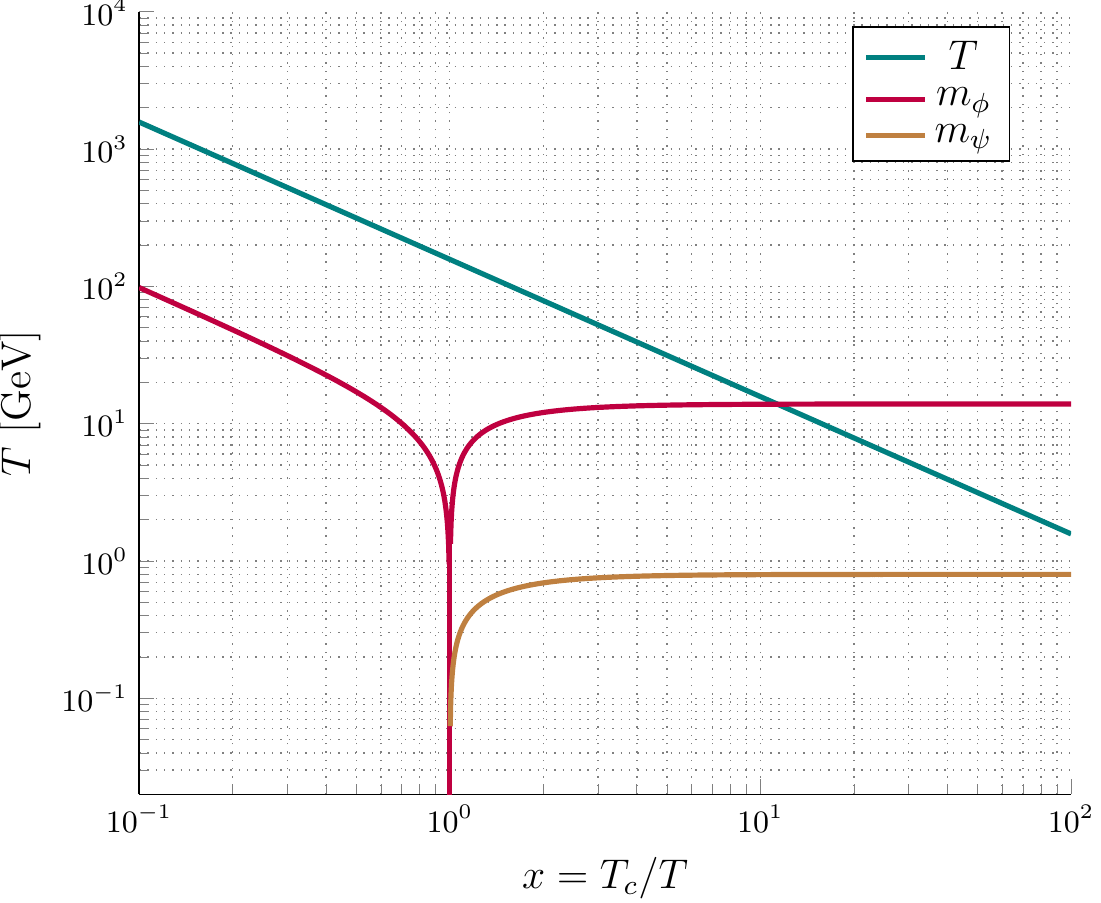}
\end{subfigure}\;\;\;\;\;\;\;\;\;\;\;\;
\begin{subfigure}[b]{0.45\linewidth}
\caption{\label{fig:b}}
\includegraphics[width=\linewidth]{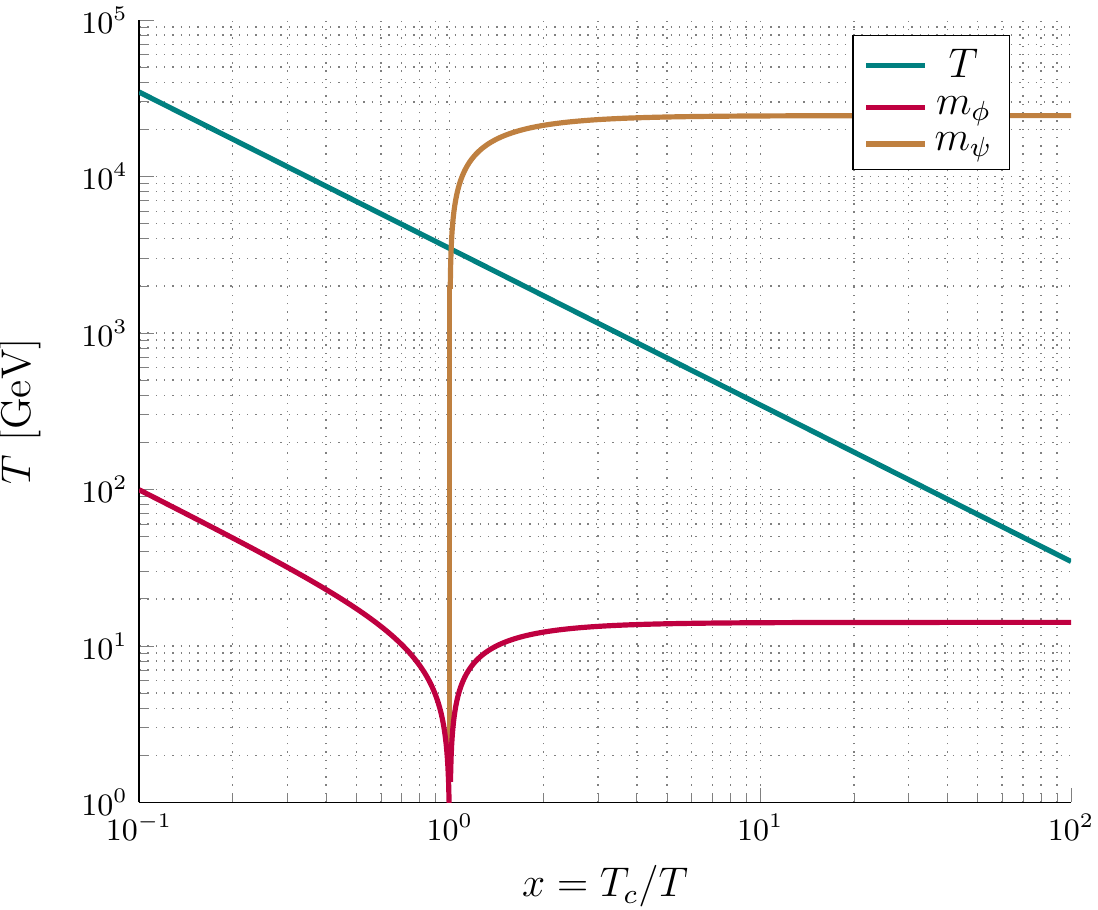}
\end{subfigure}
\begin{center}
\caption{
\footnotesize Examples reproducing (a) the usual freeze out and (b) spontaneous freeze-out scenarios in the case where $y=10^{-2}$, $\mu=10~\mathrm{GeV}$ and, respectively, $\lambda=10^{3}y^2=0.1$ and  $\lambda=10^{-2}y^4=10^{-10}$.}
\end{center}
\end{figure*} 
shows the fermion mass $m_\psi$ (brown), scalar mass $m_\phi$ (red) and temperature (teal) as functions of $x$, in the illustrative case where { $\mu=10\, \mathrm{GeV}$, $y=10^{-2}$, $n_F=2$ and $\lambda= 10^3  y^2=0.1$}. Because $x_{\rm FO}\gg \kappa$,  the temperature reaches $m_\psi$ when the mass of $\psi$ is already varying slowly, which corresponds to the usual thermal freeze-out paradigm.

\item  When $x_{\rm FO}\lesssim \kappa$,  the freeze out takes place soon after the fermionic dark-matter particles start becoming massive. At this epoch, the evolutions of the masses are
\be
\begin{aligned}
\quad \quad \quad&1\leqslant x\leqslant x_{\rm FO} :\quad m_{\phi}(x) \simeq m_\phi^{\rm tree} \sqrt{1-{1\over x^2}}\,, \\
&m_{\psi}(x) \simeq  y \mu\sqrt{{6\over \lambda+{3n_F\over 4\pi^2}y^4\log x}}\,\sqrt{1-{1\over x^2}} \,.
\end{aligned}
\ee
Because shortly after the phase transition the derivative $dm_\psi/dx$ is very large,  it is expected that the mass of the DM particles $\psi$ when they free out differs significantly from its final value today. In FIG.~\ref{fig:b}, we show the evolutions of the temperature and DM particle masses with respect to $x$ in this case. {For $\mu=10~\mathrm{GeV}$, $y=10^{-2}$, $n_F=2$ and $\lambda= 10^{-2} y^4=10^{-10}$,} one can see that $m_\psi$ becomes larger than the temperature soon after the phase transition takes place, suggesting that the DM particles $\psi$ might indeed decouple while their mass still varies significantly. This illustrates the case we are mostly interested in this work, and that we refer to as ``Spontaneous Freeze-Out''.  
\end{itemize}


 To summarize, we may have two different freeze-out scenarios
\be
\begin{aligned}\label{eq:summary}
x_{\rm FO}\gg\kappa:\quad &\text{Constant-Mass Freeze Out\,,}\\
x_{\rm FO}\simeq \kappa: \quad &\text{Spontaneous Freeze Out\,.}
\end{aligned}
\ee
Notice that in the SFO case, we discard the possibility of having $1\lesssim x_{\rm FO}< \kappa$ since, as we will see in the next section, demanding the existence of a stable minimum of the potential at late time, when the temperature vanishes, imposes a lower bound of order $\kappa$ on $x_{\rm FO}$. In practice, this condition will lead to a lower bound of order $0.03\,n_F$ on $\lambda/y^4$  (see Eq.~(\ref{eq:bound})).

\section{Dynamics of the Scalar Field \boldmath $\phi$ After Freeze Out of $\psi$}
\label{sec:scalar}

In the present section, we are interested in the late time dynamics of the scalar $\phi$, when the temperature is well below its zero-temperature mass, and its effective potential can be approximated by its zero-temperature expression.   
Because the mass of the non-relativistic DM particles $\psi$ of the relic density is $\phi$-dependent, the dust also sources the scalar potential in a way to be figured out. 
In total, the equation of motion of the scalar $\phi$ can be written as 
\be
\ddot \phi+(\Gamma_{\phi}+3H)\dot \phi=-{d\V_{\rm eff}\over d\phi}-{d\V_{\rm dust}\over d\phi}\, , 
\label{eqphi}
\ee
where the source $\V_{\rm dust}$ arises from the dust, while the contribution $\V_{\rm eff}$ is the zero-temperature effective potential of $\phi$. Note that we also include the friction term arising from the decay of $\phi$ into lighter fields. At 1-loop, the potential $\V_{\rm eff}$  involves   the Coleman-Weinberg and counterterm contribution of Eq.~(\ref{CW1}),
\be
\V_{\rm eff}(\phi)=\V_{\rm tree}(\phi)+\V_{\rm CW}(\phi)+\V_{\rm dark}^{\rm c.t.}(\phi)\,.
\label{Vfo}
\ee

In order to find $\V_{\rm dust}$, a  possible approach is to consider that beside the Einstein action coupled to the fundamental DM fields $\phi$, $\psi$, and to the SM degrees of freedom, we add an action for describing the motion of $N_\psi$ non-interacting particles of dust $\psi$. Each of them has coordinates $X^\mu_i(\tau_i)$, where $\tau_i$ is an arbitrary parameter along the world-line trajectory of particle $i$. 
Because the action of the  fields $X^\mu_i(\tau_i)$  must be independent of the parameterization, it can be chosen to be proportional to the proper time along the trajectory multiplied by the mass,  
\be
S_{\rm dust}=-\sum_i\int\! d\tau_i \,y\, |\phi(X_i)| \sqrt{g_{\mu\nu}(X_i)\, {dX^\mu_i\over d\tau_i}{dX^\nu_i\over d\tau_i}}\, ,
\ee
where $g_{\mu\nu}$ is the spacetime metric.  Varying $S_{\rm dust}$ with respect to $g_{\mu\nu}$ and $\phi$, it is possible to derive the energy density $\rho_{\rm dust}$, the pressure $P_{\rm dust}$ and the contribution $\V_{\rm dust}$ to the potential that arise from  the distribution of DM particles $\psi$~\cite{Farrar:2003uw,Damour:1994zq}, assuming that the latter is  compatible with spacetime homogeneity and isotropy of the universe at large enough scale.

However, for our purpose it is enough to multiply by $\dot\phi$ the equation of motion of $\phi$, Eq.~\eqref{eqphi}, to obtain
\be
\dot \rho_\phi+(\Gamma_\phi+3H) (\rho_\phi+P_\phi)=-\dot \phi \, {d\V_{\rm dust}\over d\phi}\, ,
\label{cphi}
\ee
where $\rho_\phi$, $P_\phi$ are the gravitational sources associated with the dark scalar,
\be
\rho_\phi = {1\over 2}\dot \phi^2+\V_{\rm eff}(\phi) \, ,\quad P_\phi = {1\over 2}\dot \phi^2-\V_{\rm eff}(\phi) \, .
\ee
Denoting $\rho_{\rm rad}$, $P_{\rm rad}$ the energy density and pressure of the visible sector, we also have~\cite{Kolb:1990vq}
\be
\dot \rho_{\rm rad}+3H (\rho_{\rm rad}+P_{\rm rad})=\Gamma_\phi (\rho_\phi+P_\phi)\,.
\label{crad}
\ee
Note that we considered here that the scalar $\phi$ decays exclusively into SM particles since we will consider in Sec.~\ref{sec:SMinteractions} the region of the parameter space where the mass of $\psi$ is always larger than the mass of $\phi$. Combining  Eqs~(\ref{cphi}),~(\ref{crad}) and the conservation of the stress-energy tensor associated with the total system composed of the scalar $\phi$, the massless SM and the dust, we obtain~\cite{Hoffman:2003ru,Franca:2003zg,Anderson:1997un,Rosenfeld:2005pw}
\be
\dot \rho_{\rm dust}+3H(\rho_{\rm dust}+P_{\rm dust})=\dot \phi\,  {d\V_{\rm dust}\over d\phi}\, .
\label{conpart}
\ee
We are interested in the case where the relic dark-matter particles $\psi$ are non-relativistic. Assuming for simplicity the limit case of vanishing particle velocities,  the DM fluid is pressureless, while its energy density originates only from invariant mass,
\be\label{eq:dust}
\rho_{\rm dust}=n_\psi y |\phi|\,, \quad P_{\rm dust}=0\, .
\ee
Using Eq.~\eqref{conpart} and the fact that $n_\psi= N_\psi/a^3$, where $a$ is the scale factor of the universe, we obtain 
\be
{d\V_{\rm dust}\over d\phi}=\mathrm{sign}(\phi)y n_\psi \,.
\ee

The effect of  dust on the dynamics of the scalar $\phi$ after the fermionic DM freezes out is however minor, due to the suppression factor $1/a^3$ in $n_\psi$. To see this qualitatively, we may neglect the Coleman-Weinberg and counterterm contributions to the scalar effective potential, and study the minima of $\V_{\rm tree}+\V_{\rm dust}$.  This potential admits a local minimum at some $\langle \phi\rangle>\mu \sqrt{2/\lambda}$ when
\be\label{eq:condition}
n_\psi< n_c={2\sqrt{2}\over 3}\, {\mu^3\over y\sqrt{\lambda}}\, . 
\ee
Thanks  to the expansion of the universe, or if the DM particle $\psi$ is sufficiently heavy (and its number density small enough), the above condition always ends to be satisfied and $\langle \phi\rangle$ can converge to the tree-level value $\langle \phi\rangle_{\rm tree}$, given in Eq.~\eqref{ttree}. Note however that in principle the back-reaction of the dust might destabilize the vacuum $\langle\phi\rangle>0$, right after the freeze out takes place. This would lead $\phi$ to decrease and possibly allow the dark-matter particles $\psi$ to re-thermalize. However, such a behavior would be contradictory with the hypothesis that dark matter is non relativistic while estimating the back-reaction. Therefore a more thorough treatment of this possibility in the presence of a quasi-relativistic velocity distribution would be required. We let such a dedicated study for future work. 

Neglecting this back-reaction of the relic density, another source of destabilization of the vacuum $\langle \phi \rangle >0$ may arise from the 
Coleman-Weinberg and counterterm contributions to the effective potential. In the SFO regime, corresponding to the limit of small self coupling $\lambda$ as compared to $y^4$, one can understand this point by neglecting the contribution to the Coleman-Weinberg potential arising from the scalar sector.  The analytic expression of the vev of $\phi$ that minimizes the potential is then  found to be
\begin{equation}
\langle\phi\rangle\simeq\frac{4\pi \mu}{y^2\sqrt{-n_F W\!\left(-\frac{16\pi^2\mu^2\exp(-1-8\pi^2\lambda/3n_F y^4)}{n_F Q^2 y^2}\right)}}\,,
\end{equation}
where $W(z)$ stands for the product logarithm (or Lambert $W$ function). Imposing the above value to be real turns out to demanding the argument of $W(z)$ to be larger than $-1/e$, which in turn leads to the condition
\begin{equation}\label{eq:bound}
\frac{\lambda}{y^4}>\frac{3 n_F}{8\pi^2}\left(\ln\frac{2}{3}+2\gamma_{\rm E}\right)\simeq 0.03\, n_F\,.
\end{equation}
In non-SFO cases, larger values of the self coupling $\lambda$ always imply the existence of a local or global minimum at some $\langle\phi\rangle_0>0$.
Therefore, as long as the bound of Eq.~\eqref{eq:bound} is satisfied, the vev of $\phi$ converges towards the minimum of the potential at zero temperature and zero back-reaction of the dust,  while the mass $m_\psi$ approaches its final constant value,
\be
m_\psi^0=y\, \langle \phi \rangle_0\, .
\ee
\section{Interaction with the Standard-Model Bath}
\label{sec:SMinteractions}

In the previous sections, we have described the dynamics of a scalar field living in a thermal potential. We have in particular assumed that the dark-matter particles $\psi$ and the dark scalar $\phi$ were maintained in thermal equilibrium with the Standard Model before the spontaneous freeze-out mechanism takes place.
However, we have not specified yet in which way the dark sector interacts with Standard-Model particles. In this section we explicitly introduce such interactions, perform a numerical scan and derive important constraints on the parameter space.

Although one could consider a complete set of dimension-six operators for describing the annihilation of the dark-matter particles $\psi$ into SM fermions,  we will consider in this paper only spin-independent  interactions as an educational toy example and let a more systematic study for future work. Therefore, we focus on the two simplest  operators of this kind leading to $s$-wave and $p$-wave processes for dark-matter annihilation into SM\footnote{Note that in principle the fermion $f$ may not be part of the Standard-Model spectrum, as long as its interactions with it are sufficient to keep $f$  in equilibrium with the SM bath until dark-matter freezes out.} fermions $f$, that are respectively through vector (V) and scalar (S) operators 
\begin{equation}\label{eq:operators}
\mathcal O_V = \bar \psi {\gamma_\mu}\psi \bar f {\gamma^\mu}f\quad\text{and}\quad \mathcal O_S =\bar \psi\psi \bar f f\,.
\end{equation}
Introducing the couplings $G_V$ and $G_S$, one can write the associated cross-sections of dark-matter annihiliation $\psi\bar \psi\to f\bar f$ as 
\be
\begin{aligned}
\sigma_V &=\frac{G_V^2}{32 \pi } \sqrt{\frac{s-4 m_f^2}{s-4 m_\psi^2}} \\
&\times \bigg(\frac{(s-4 m_\psi^2) (s-4 m_f^2)}{3 s}+4 \left(m_\psi^2+m_f^2\right)+s\bigg),\\ 
\sigma_S &= \frac{G_S^2}{32 \pi \, s} \sqrt{\frac{s-4 m_f^2}{s-4 m_\psi^2}} \left(s-4 m_\psi^2\right)\! \left(s-4 m_f^2\right).
\end{aligned}
\ee
Assuming that the mass of the fermion $f$ is much lighter than the dark-matter particle in the range $1<x\leqslant x_{\rm FO}$, and using the fact that  the relative velocity of thermalized dark-matter particles satisfies $\langle v^2\rangle = 6 T/m_\psi$, one can express the thermally-averaged cross-section of annihilation as
\be
\begin{aligned}\label{eq:thermalXsection}
\langle \sigma v\rangle_V &\simeq  \frac{G_V^2}{2 \pi }\Big(1+\frac{x^{-1}T_c}{m_\psi(x)}\Big)\,m_\psi^2(x)\,,\\
\langle \sigma v\rangle_S&\simeq  \frac{3G_S^2 }{8 \pi }\, x^{-1} T_c m_\psi(x)\,. 
\end{aligned}
\ee
\subsection{Relic Density}
The evolution of the dark-matter number density $n_\psi$ can be described by the Boltzmann equation, which we express in terms of the yield $Y_\psi=n_\psi/s$,
\begin{equation}
\frac{\dd Y_\psi}{dx}=\frac{\langle \sigma v\rangle s}{x H}(Y_{\psi,\rm{eq}}^{2}-Y_\psi^2)\,.
\end{equation}
In this equation, $s$ is the entropy density of the universe and $Y_{\psi,\rm{eq}}=n_{\psi,\rm{eq}}/s$ is the value of the yield when $\psi$ follows a Boltzmann equilibrium. After dark-matter particles and anti-particles freeze out, the relic density of dark matter is given by
\begin{equation}
\Omega h^2 =n_F\frac{m_\psi(x_0)s_0}{6H_0^2 M_{\rm P}^2}\, Y_{\psi}^0\,,
\end{equation}
where indices ``0" express the fact that the different quantities are evaluated at the present time, where $\mbox{$T=T_0\simeq 2.4\times 10^{-4}$\, eV}$. The entropy density of the universe is therefore given by $s_0=2.1\times 10^{-38}\,\mathrm{GeV^3}$, the Hubble parameter is $H_0\simeq  71\,\mathrm{km.s^{-1}.Mpc^{-1}}$, $h\equiv H_0/(100\,  \mathrm{km.s^{-1}.Mpc^{-1}})$ and the Planck mass is denoted by $M_{\rm P}=1.22 \times 10^{19}$~GeV.


\subsection{Numerical Results}

Before presenting our numerical results, let us make a few comments on what is expected in our scenario of spontaneous freeze out, as compared to the usual case of a constant-mass dark matter particle.
\begin{itemize}
\item For a given dark-matter mass $m_\psi^0$ and relic density at present time, the dark-matter yield at freeze out $Y_\psi^{\rm FO}$ is fixed.
\item The dark-matter relative velocity at freeze out  \mbox{$\langle v^2\rangle\sim T_{\rm FO}/m_\psi=\kappa^{-1}$} is essentially model independent. Because in our scenario $m_\psi(x_{\rm FO})<m_\psi^0$, we expect the freeze-out temperature in the SFO case to be lower than in the constant-mass standard WIMP scenario.
\item At freeze out, the condition $n_\psi^{\rm FO}\langle\sigma v\rangle_{\rm FO} =H_{\rm FO}$ can be expressed as 
\begin{equation}
Y_\psi^{\rm FO}\langle\sigma v\rangle_{\rm FO}\propto T_{\rm FO}^{-1}\,.
\end{equation} 
Therefore the dark-matter annihilation cross section at freeze out $\sigma_{\rm FO}$ in the SFO case is larger than in the constant-mass paradigm.
\item In our benchmark models, the cross sections of Eq.~\eqref{eq:thermalXsection} evolve after freeze out as
\be
\begin{aligned}
\quad \langle \sigma v\rangle_V &\simeq  \frac{G_V^2}{2 \pi }\Big(1+\frac{v^2}{6}\Big)\,m_\psi^2(x)\,,\\
\quad \langle \sigma v\rangle_S&\simeq  \frac{3G_S^2 }{48 \pi }\, v^2 m_\psi(x)\,.
\end{aligned}
\ee
Because in the SFO case the dark-matter particle mass increases with time,
the ratio between the annihilation cross section in our scenario as compared to the usual WIMP case increases from the time of freeze out to present time, where the DM velocity $v\simeq 200~\mathrm{km.s^{-1}}$ is model independent.
\end{itemize}

\begin{figure*}
\begin{center}
\includegraphics[width=0.8\linewidth]{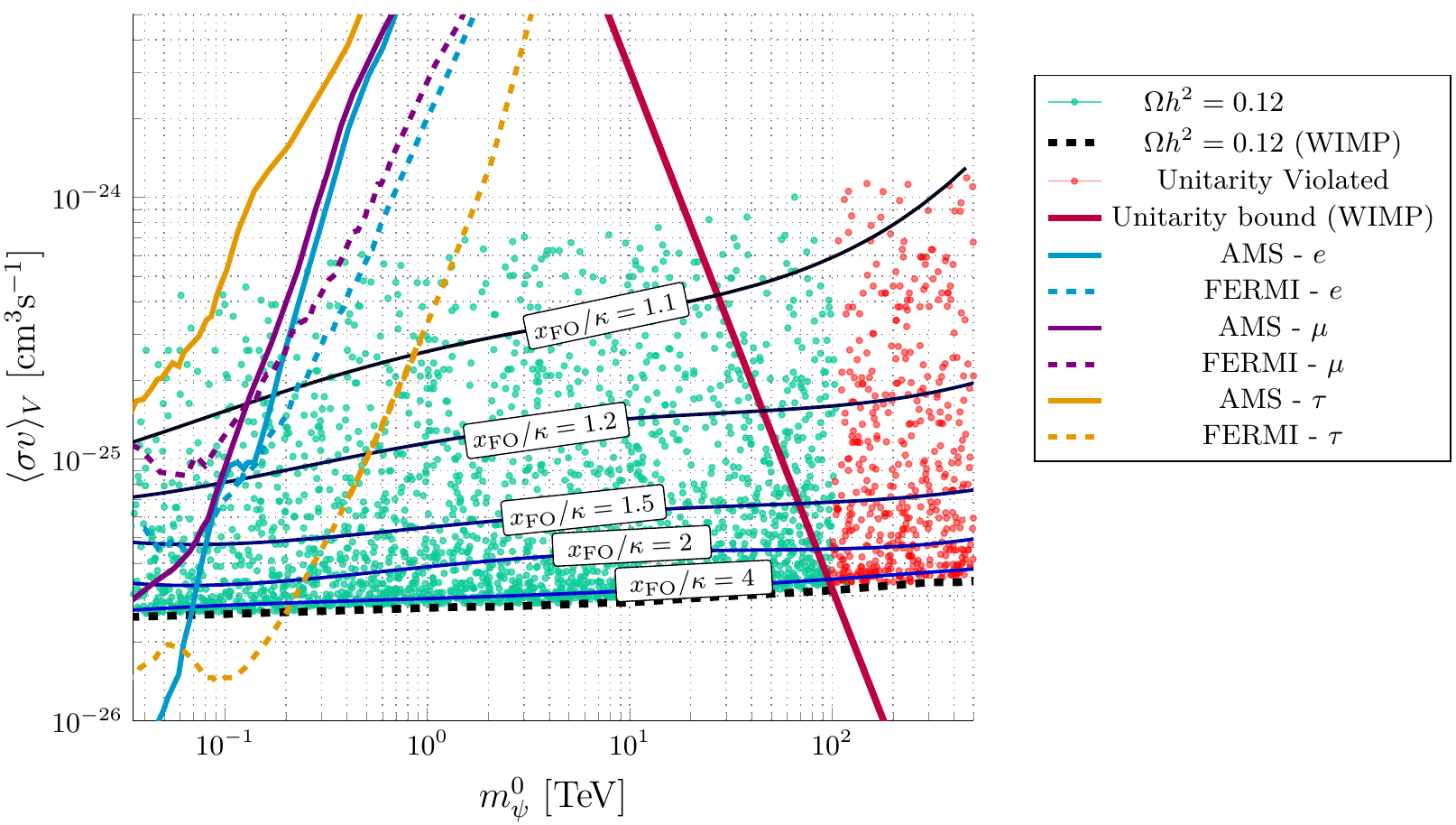}
\end{center}
\caption{\label{fig:IDswave}  \footnotesize Numerical results for the cross section of annihilation of dark-matter particles interacting with SM fermions via the operator $\mathcal{O}_V$. The plain red line indicates the standard WIMP unitarity bound, wheareas the red dots stand for the points which violate unitarity in our scenario. Indirect-detection constraints on the annihilation cross-section of a dark-matter candidate interacting with one single species of lepton are indicated.}
\end{figure*}
\begin{figure*}
\begin{center}
\includegraphics[width=0.8\linewidth]{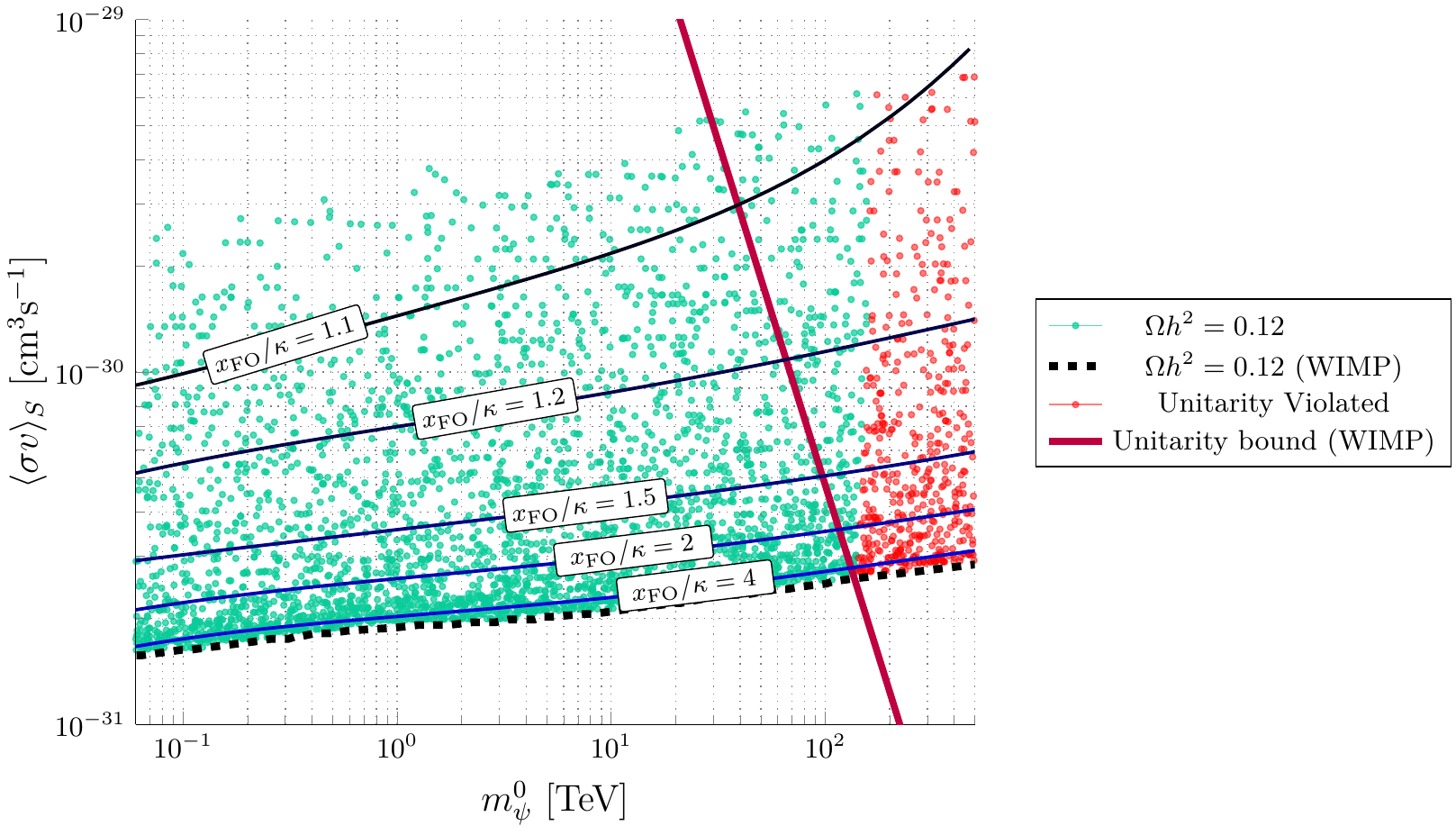}
\end{center}
\caption{\label{fig:IDpwave}  \footnotesize Numerical results for the cross section of annihilation of dark-matter particles interacting with SM fermions via the operator $\mathcal{O}_S$. The plain red line indicates the standard WIMP unitarity bound, wheareas the red dots stand for the points which violate unitarity in our scenario.}
\end{figure*}
In Figs~\ref{fig:IDswave} and \ref{fig:IDpwave} we present our numerical results for the two benchmark operators $\mathcal{O}_V$ and $\mathcal{O}_S$. In both figures, we scan over the whole parameter space $\{\mu,\lambda,y,G_{V,S}\}$ and compute the annihilation cross section necessary to obtain the correct relic abundance of dark matter \mbox{$\Omega h^2\simeq 0.12$}. In both cases our findings are very similar and in  very good agreement with the qualitative predictions that we have just developed about our SFO scenario. Indeed one can clearly see that data points which lead to low values of the ratio $x_{\rm FO}/\kappa$, and therefore correspond to the SFO case (as defined in Eq.~\eqref{eq:summary}) show an annihilation cross section which can be significantly larger than in the usual freeze-out scenario by more than one order of magnitude. As a matter of fact, because the ratio $\lambda/y^4$ admits a lower bound given in Eq.~(\ref{eq:bound}),  it can be seen from Eq.~\eqref{eq:xFO} that the ratio $x_{\rm FO}/\kappa$ cannot be arbitrarily small. Therefore, the increase of the annihilation cross-section  cannot be arbitrarily large. 

Another interesting feature of our scenario is that a fraction of the parameter space which is ruled out by the unitarity constraint in the constant-mass WIMP scenario remains allowed in the SFO case. Indeed, the unitarity bound that we obtain is now independent on the cross section of annihilation today and  amounts to an upper bound on the DM mass which is unchanged as compared to the constant-mass case.

\begin{figure*}
\includegraphics[width=0.485\linewidth]{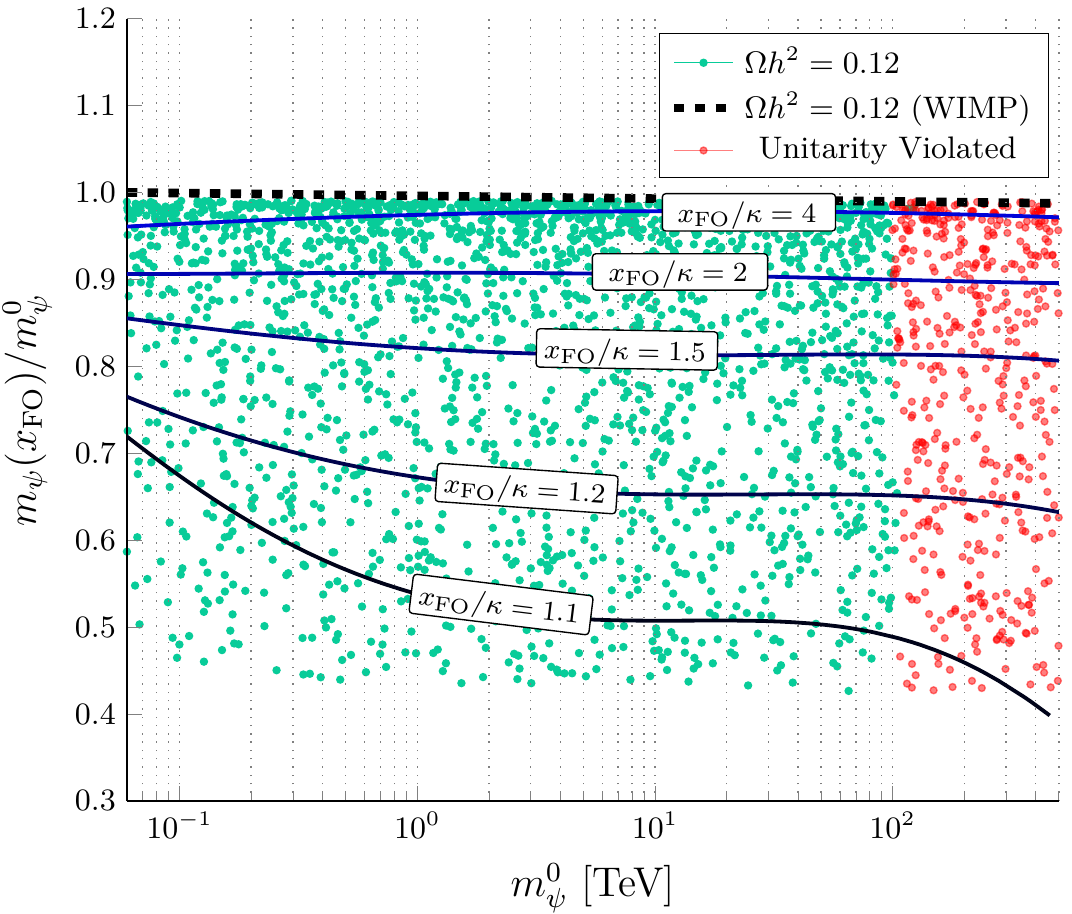}\hspace{0.02\linewidth}\includegraphics[width=0.485\linewidth]{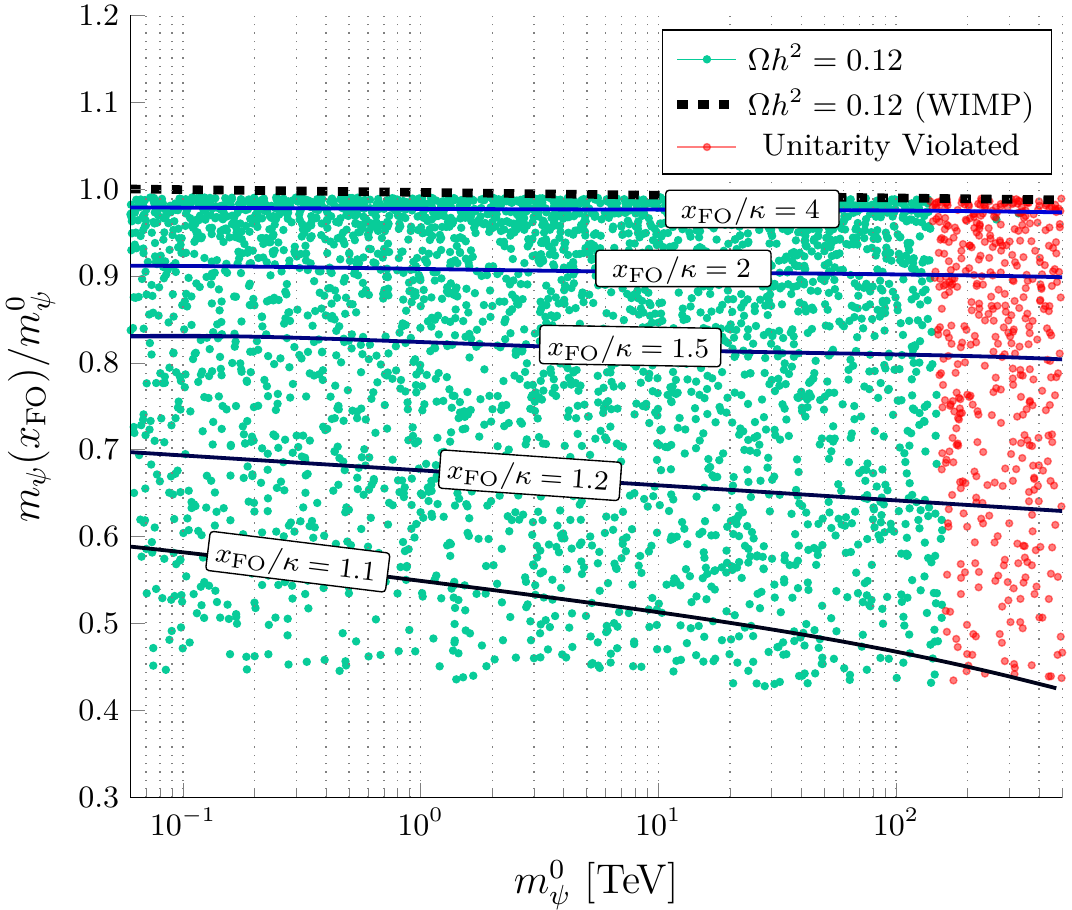}
\caption{\label{fig:Ratios}\footnotesize Ratio between the values of the dark-matter particle mass at the time of freeze-out and at present time for data points leading to the correct relic abundance of dark matter. The left panel corresponds to an $s$-wave annihilation cross section (operator $\mathcal O_V$), whereas the right panel stands for the $p$-wave annihilation cross section (operator $\mathcal O_S$). Red circles are associated with data points for which the annihilation cross sections violate unitarity at the time of freeze out.}
\end{figure*}

In Fig.~\ref{fig:Ratios} we also present the value of the ratio between the dark-matter particle mass at the time of freeze out and its value at present time. 

\subsection{Phenomenological Constraints}

As a direct consequence of such an increased cross section between the dark sector and the visible sector, models of thermal dark matter which were already experimentally ruled out in the usual constant-mass freeze out scenario turn out to be even more excluded in the SFO regime.

As a result, a model in which a dark-matter particle interacting with coloured fermions via the vectorial effective operator $\mathcal{O}_V$ is totally ruled out by direct-detection constraints in our setup. Therefore, the only fermions with whom DM particles can interact in this case are charged leptons or neutrinos. Although the interaction of DM particles with electrons could induce some electron recoil, future direct-detection experiments are still far from constraining the region of parameter space which is of interest in our scenario. 

In the case where DM interacts with coloured particles through the scalar operator $\mathcal{O}_S$,  direct-detection experiments have ruled out a significant fraction of the parameter space, but heavy DM candidates could still, in principle, escape detection for masses above  $\mathcal{O}(1)$TeV. However, as we will see in the next subsection, the existence of such an operator leads to an effective Yukawa coupling between the dark scalar $\phi$ and SM fermions. While $\phi$ acquires a large vev, such coupling can contribute dangerously to the mass of these fermions. As a matter of fact, it turns out that demanding such corrections to not overshoot the current experimental uncertainties on the quark masses eliminates all the data points which could escape the current direct detection limits set by Xenon-1T~\cite{Aprile:2019dbj}. Therefore we are to consider only in what follows that dark-matter particles interact with SM leptons.

\subsubsection*{Indirect Detection}

In the case of a velocity-suppressed cross-section ($p$-wave), the annihilation of dark-matter particles in the galaxy today cannot lead to any visible signal under the form of cosmic rays, unless enhanced by long-range interactions (Sommerfeld enhancement). However, in the case of an $s$-wave annihilation process, there exists a plethora of constraints which can be used to set stringent limits on our parameter space. {Indeed, for masses $\lesssim 10~\rm{GeV}$, the annihilation of dark-matter particles, even after they have frozen out, injects a fraction of energy which can distort significantly the cosmic microwave background~\cite{Slatyer:2015jla}, leading to severe constraints on the annihilation cross-section of dark-matter particles.} Moreover, above  10 GeV, the FERMI collaboration measuring the spectrum of Dwarf Spheroidal  Galaxies  in  the  Milky  Way \cite{Ackermann:2015zua, Fermi-LAT:2016uux},  and  the  Alpha Magnetic Spectrometer (AMS) detecting cosmic-rays  set the  most  robust  limits on dark-matter annihilation in the galaxy \cite{Aguilar:2014mma,Accardo:2014lma}. 
In Fig.~\ref{fig:IDswave}, we indicate the different constraints which have been derived in Ref.~\cite{Leane:2018kjk} for various possible leptonic final states.  The fact that the annihilation cross section in the SFO case is significantly larger than in the usual constant-mass scenario leads to much more severe constraints. Therefore, heavy dark-matter candidates, which are unlikely to be detected in the near future in the usual constant-mass WIMP scenario, turn out to be very soon accessible to collaborations such as FERMI or AMS in the SFO case.


\subsection{Constraints on the Scalar Sector}

\begin{figure}[b]

  \vspace{0cm}
\begin{tikzpicture}[ thick    ]

\draw[scalar1] (-1.25,0)node[left]{$\phi$}--(0.5,0);

 \draw[electron] (2.5,0) arc (0:180:1) ;
 \draw[antielectron] (2.5,0) arc (0:-180:1) ;
 
 \draw[dark] (2.5,0)--(4.5,1.)node[right]{$f$} ;
 \draw[antidark] (2.5,0)--(4.5,-1.)node[right]{$\bar f$} ;

\draw (1.3,1.3)node[right]{$\bar\psi$};
\draw (1.2,-1.3)node[right]{$\psi$};
  \shade[top color=viol, bottom color=white]
 (0.5,0)circle(0.12);
 \shade[top color=greeen, bottom color=white]
 (2.5,0)circle(0.12);

 \draw (0.2,-0.4)node{$y$};
   \draw (2.8,-0.5)node{$G_{S}$};

 \end{tikzpicture}
 \caption{\label{fig:decay} \footnotesize Decay of the scalar $\phi$ into SM fermions via a loop of dark-matter particles.}
\end{figure}
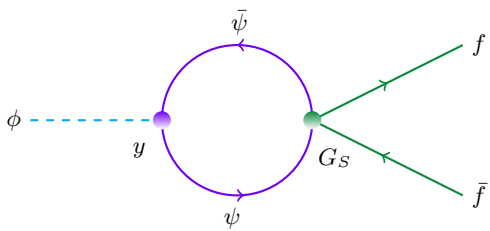

As we have anticipated in Sec.~\ref{sec:FO}, throughout the universe history, the scalar field $\phi$ can have access to different decay channels: For $x<1$, we have seen that the fermionic dark-matter particles are massless and $\phi$ can decay at tree level, with the lifetime
\begin{equation}\label{eq:decay}
x<1\,:\quad(\tau_\phi)^{-1}=\Gamma_{\phi\to\bar\psi\psi}=\frac{y^2}{8\pi}m_\phi(x)\,.
\end{equation}
After the phase transition, when $\lambda\ll n_F y^2$, whether we are in the spontaneous freeze out case  (where this condition is automatically satisfied) or not, the ratio of the masses in the dark sector can be expressed as follows,
\be
\label{eq:ratio}
1\leqslant x\leqslant x_{\rm FO} :\quad \frac{m_\psi(x)^2}{m_\phi(x)^2}\simeq\frac{3y^2}{\lambda+\frac{3n_F}{4\pi^2}y^4\log x}\gg 1\,.
\ee
As a result, in this region of parameter space, the decay and inverse-decay $\phi\leftrightarrow \psi+\bar \psi$ are now kinematically forbidden. However, at the loop level, certain opertors can lead to an effective decay of the dark scalar into SM particles, as depicted in Fig.~\ref{fig:decay}. 
In the case of the scalar operator $\mathcal O_S$, the effective coupling induced by such a diagram is
\be
\begin{aligned}\label{eq:effectiveop}
\mathcal O_S^{\rm eff}&\sim\left(y\frac{G_S m_\psi^2}{16\pi^2}\right)\phi\bar f f\,,
\end{aligned}
\ee
and it provides the scalar  lifetime (for $x> 1$)
\be
\begin{aligned}
(\tau_\phi^S)^{-1}&=\Gamma_\phi^S\sim\frac{m_\phi}{8\pi} \left(\frac{ y\, G_{S} m_\psi^2}{16 \pi^2}\right)^2\left(1-\frac{4 m_f^2}{m_\phi^2}\right)^{3/2}\,.\\
\end{aligned}
\ee
It is important to notice that the presence of the effective Yukawa coupling of Eq.~\eqref{eq:effectiveop} between the dark scalar $\phi$ and SM fermions can source a correction to the mass term of the latter after $\phi$ acquires a large vev. Since the dark-matter mass is given by $m_\psi= y\langle\phi\rangle$, it is clear from Eq.~\eqref{eq:effectiveop} that the corresponding correction to the mass of the SM fermion with whom the dark-matter particles $\psi$ interact, is of order
\begin{equation}
\Delta m_f \sim G_S{m_\psi^3}\,.
\end{equation}
In practice, given the relatively large value of the coupling $G_S$ in the SFO case, we obtain a contribution to the mass of the SM fermion $f$ of order \mbox{$\Delta m_f=  \mathcal O(10^{-3})\, m_\psi$}. Given the extremely good accuracy with which the SM fermion masses are measured, the presence of such contribution imposes an upper bound on the dark-matter particle mass which depends on the details of the model considered.

However, in the case of the vectorial operator $\mathcal O_V$ such an effective coupling is forbidden. Therefore, unless adding contact interactions between $\phi$ and SM particles, the dark scalar could in principle be relatively stable on cosmological scales. The possibility that the dark scalar would be long-lived after the phase transition takes place can have four major consequences: 
\begin{itemize}
\item[$(i)$] If the lifetime of the scalar field $\phi$ is longer than the age of the universe, it could participate to the DM relic abundance together with the $\psi$ particles after freezing out from the thermal bath.
\item[$(ii)$]  The coherent oscillations of $\langle \phi\rangle$ around the minimum of the potential  when the temperature of the universe has dropped below the mass of $\phi$ might contribute significantly to the matter abundance and overclose the universe~\cite{Preskill:1982cy,Abbott:1982af,Dine:1982ah}. This problem, which is known as the Polonyi problem~\cite{Coughlan:1983ci,Ellis:1986zt}, has in particular been raised in the context of Supergravity and String Theory~\cite{deCarlos:1993wie}.
\item[$(iii)$]  If the energy density of the scalar (either due to its coherent oscillations or to its relic density) comes to dominate the energy density of the universe before the scalar decays into SM particles, the corresponding entropy injection into the visible bath is known to reduce the dark-matter relic abundance \cite{Kolb:1990vq,Heurtier:2019eou,Berlin:2016gtr}. In that case, this effect has to be taken into account in the numerics. 
\item[$(iv)$] Finally, if $\phi$ decays after the Big-Bang Nucleosynthesis takes place, it can destroy the predictions for the different atom density fractions in the universe today.
\end{itemize}

For these different reasons, it is essential that the dark scalar possesses a significant decay width after the phase transition takes place. Therefore we assumed in our scenario that such a decay rate is present and does not lead the dark scalar to contribute to the relic abundance at all nor to dominate the energy density of the universe at any time.

As a matter of fact, whether the effective operator of Eq.~\eqref{eq:effectiveop} exists or not for a given interaction of dark-matter particles $\psi$ with the SM, it is expected that our dark scalar can mix with the SM scalar. Therefore, depending on the mixing of $\phi$ with the Higgs boson, the former can possess a significant decay width after the phase transition happens. This in turn can solve all the issues listed above. Note that in the case of the scalar operator we introduced in Eq.~\eqref{eq:effectiveop}, a mixing of the scalar $\phi$ with the Higgs boson arises at the 2-loop level. In any case, a particular attention should be given to the scalar mass matrix and how the dark phase transition might affect the Higgs sector.

\subsection{Effect of the Non-Adiabaticity}

Up to now, we have been assuming that the scalar field $\phi$ systematically tracks the minimum of its thermal potential in an adiabatic fashion. We have seen that such a tracking can force the freeze-out temperature of dark matter to be lower than what one would infer from the usual constant-mass WIMP scenario. We showed that this lowering of the freeze-out temperature leads to a significantly larger annihilation cross-section of dark matter in order to accomodate the correct relic abundance.

\begin{figure}
\includegraphics[width=\linewidth]{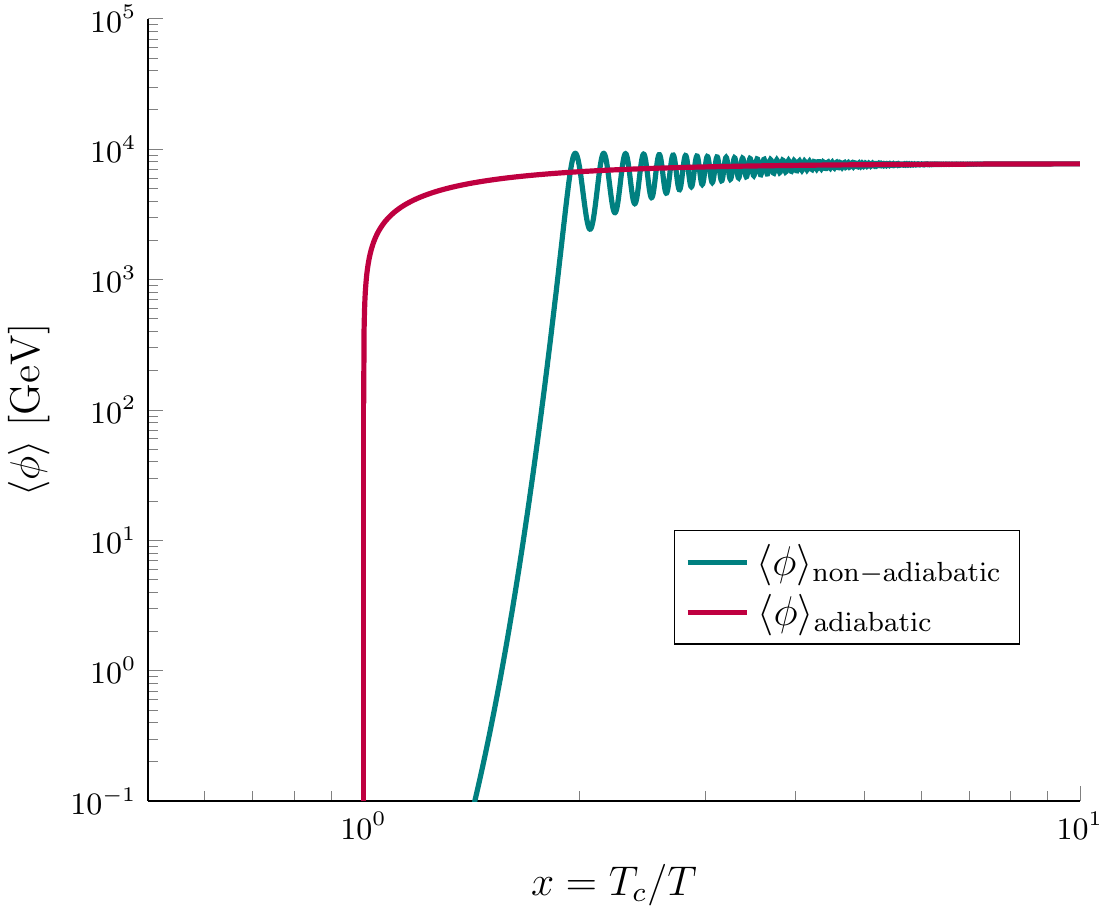}
\caption{\label{fig:stabilization} \footnotesize Numerical simulation of the scalar field dynamics after the phase transition takes place for $y=10^{-2}$, $\lambda=10 y^4=10^{-7}$, $\mu=1$~GeV and including the scalar decay width $\Gamma_\phi=10^{-2}\mu$.}
\end{figure}

In practice, when the phase transition takes place, the thermal mass of the scalar field is close to zero, and it takes time before $\langle\phi\rangle$ falls and oscillates around the minimum of its potential. One can visualize this effect in Fig.~\ref{fig:stabilization} on a particular example. This leads the dark-matter mass to overshoot the universe temperature at an even later time (i.e. lower temperature) than what we have been using in our numerical simulations. Therefore, the true value of $m_\psi(x_{\rm FO})$ is expected to be lower than that we have described. In other words, the analysis that we performed throughout this work is relatively conservative and is likely to underestimate the effect that the SFO regime can have on the annihilation cross section of dark-matter particles, as well as on the increase of the DM mass between the freeze-out time and today. Although we restrict ourself to an adiabatic description of the SFO mechanism in this work, we let the thorough study of this effect in the presence of a non-adiabatic behavior of the scalar for future work.

\section{Conclusion}\label{sec:conclusion}

In this paper, we have focused on the simple case in which a dark-matter fermionic particle acquires its mass from the spontaneous breaking of a global $\mathbb Z_2$ symmetry, and is in thermal equilibrium with the Standard-Model bath. We have computed the thermal corrections to the scalar potential which arise from the contribution of thermalized dark-sector particles to the free energy. The spontaneous breaking of the global $\mathbb Z_2$ symmetry is driven by a second order phase transition, similarly to what happens to the Higgs boson in the early universe during the electroweak phase transition. We have demonstrated that the dynamics of such a phase transition, through which dark-matter particles spontaneously acquire mass, might interfere with the thermal freeze-out mechanism. We have in particular identified an interesting region of the parameter space where the vacuum expectation value of the scalar field is large enough for the second-order phase transition to enforce a {\em spontaneous freeze out} of dark-matter particles from the thermal bath. We have studied the phenomenology of our model in this regime in the simple case where dark-matter particles  interact with SM fermions through spin-independent dimension-six operators. We have scanned over the parameter space in order to accomodate the relic-density constraint, and we have compared our results to the most recent limits on dark-matter direct and indirect detection from XENON 1T, FERMI and AMS, respectively depending on whether DM particles interact with coloured particles or leptons.

To put it in a nutshell, we have shown that the SFO regime enforces DM particles to decouple at a lower freeze-out temperature than in the usual constant-mass models of WIMPs. Moreover, in the cases we have studied, the SFO regime favors an annihilation cross section of DM particles into SM states that is larger by more than one order of magnitude than that required in the WIMP paradigm, in order to obtain the correct relic abundance. This renders our model of SFO more experimentally testable than the constant-mass WIMP scenario. Last but not least, we have discussed the fact that the presence of non-adiabaticity in the way the dark scalar follows the minimum of its potential typically leads to an enhancement of the effect we have predicted.

As a conclusion,  we would like to emphasize that in every model of thermal dark matter in which the dark-matter particle mass is generated via spontaneous breaking of a high-energy global symmetry, one should in principle pay attention to the fact that thermal corrections to the scalar potential might affect significantly the freeze-out mechanism. For a global $\mathbb Z_2$ symmetry breaking, we have shown that such corrections become significant if the quartic coupling $\lambda$ and the Yukawa coupling $y$ lay in the regime $\lambda\lesssim y^4$.


\section*{Acknowledgement}

The authors would like to thank D. Chowdhury, E. Dudas, T. Hambye,  F. Huang, K. Kaneta, D. Kim, F. Kling, M. Pierre, U. Reinosa and  D. Teresi for lively discussions and useful inputs during the realization of this work. L.H. would also like to thank G.-A. Soto for  inspiring discussions throughout the realization of this work.
The  research  activities  of  L.H. are supported in part by the Department of Energy under Grant DE-FG02-13ER41976 (de-sc0009913). This work was partially performed at the Aspen Center for Physics, which is supported by National Science Foundation grant PHY-1607611, as well as at the CERN Theory Department. This work was made possible by Institut Pascal at Universit\'e
Paris-Saclay with the support of the P2I and SPU research departments and
the P2IO Laboratory of Excellence (program “Investissements d’avenir”
ANR-11-IDEX-0003-01 Paris-Saclay and ANR-10-LABX-0038), as well as the
IPhT.  

\bibliography{Draft}
\bibliographystyle{apsrev4-1}

  
\end{document}